\documentclass{article}
\usepackage{waspaa19,amsmath,graphicx,url,times}
\usepackage{color}
\usepackage{booktabs}
\usepackage{microtype}
\usepackage{multirow}
\usepackage{siunitx}
\usepackage[T1]{fontenc}

\title{Cross-task learning for audio tagging, sound event detection and spatial localization: DCASE 2019 baseline systems}

\name{Qiuqiang Kong$^1$, Yin Cao$^1$, Turab Iqbal$^1$, Yong Xu$^2$, Wenwu Wang$^1$, Mark D. Plumbley$^1$}
\address{$^1$Centre for Vision, Speech and Signal Processing (CVSSP), University of Surrey, UK \\
\{q.kong, yin.cao, t.iqbal, w.wang, m.plumbley\}@surrey.ac.uk \\
$^2$Tencent AI lab, Bellevue, USA \\
lucayongxu@tencent.com
}

\begin{document}

\ninept
\maketitle

\begin{sloppy}

\begin{abstract}
  The Detection and Classification of Acoustic Scenes and Events (DCASE) 2019 challenge focuses on audio tagging, sound event detection and spatial localisation. DCASE 2019 consists of five tasks: 1) acoustic scene classification, 2) audio tagging with noisy labels and minimal supervision, 3) sound event localisation and detection, 4) sound event detection in domestic environments, and 5) urban sound tagging. In this paper, we propose generic cross-task baseline systems based on convolutional neural networks (CNNs). The motivation is to investigate the performance of a variety of models across several audio recognition tasks without exploiting the specific characteristics of the tasks. We looked at CNNs with 5, 9, and 13 layers, and found that the optimal architecture is task-dependent. For the systems we considered, we found that the 9-layer CNN with average pooling after convolutional layers is a good model for a majority of the DCASE 2019 tasks.
\end{abstract}


\section{Introduction}
\label{sec:intro}

Sound carries a large amount of information that can be utilised in a number of applications, such as information retrieval \cite{turnbull2008semantic}, abnormal event detection \cite{clavel2005events} and autonomous cars \cite{xu2018large}. The Detection and Classification of Acoustic Scenes and Events (DCASE) is a series of challenges aimed at developing sound classification and detection systems. The first DCASE challenge was organized in conjunction with WASPAA 2013 \cite{giannoulis2013detection}, and is now an annual challenge and workshop \cite{mesaros2017dcase, mesaros2018multi, dcase2019_link}. The most recent DCASE challenge is the DCASE 2019 \cite{dcase2019_link}. In this paper, we propose a number of cross-task baseline systems for the DCASE 2019 challenge. These systems are designed to be applicable to a wide variety of tasks. 

Previous work in DCASE challenges mainly focus on particular tasks \cite{mesaros2016tut, mesaros2018multi}. That is, the recognition systems are built for individual tasks. However, a system trained on one task may not reflect its generalisation ability on other tasks. The motivation of this paper is to investigate how generic systems perform across several task. We call such systems \textit{cross-task} systems. As convolutional neural networks (CNNs) have achieved state-of-the-art performance in many DCASE challenge tasks \cite{mesaros2016tut, mesaros2017dcase, mesaros2018multi}, we investigate different CNN architectures as possible cross-task systems. 

The DCASE 2019 challenge consists of five tasks. In Task 1, \textit{Acoustic scene classification (ASC)} \cite{mesaros2018multi}, the aim is to classify audio recordings, recorded in a public area, into one of several predefined acoustic scene classes. This task includes three subtasks: a matching device ASC subtask, a mismatching device ASC subtask and an open set ASC subtask. In Task 2, \textit{Audio tagging with noisy labels and minimal supervision} \cite{task2_link}, the aim is to predict the tags of audio recordings while utilising a small number of manually-verified labels and a much larger number of noisy labels. In Task 3, \textit{Sound event localization and detection (SELD)} \cite{adavanne2018sound}, the aim is to predict the presence or absence of sound events, their onset and offset times, and their spatial locations in azimuth and elevation. In Task 4, \textit{Sound event detection in domestic environments} \cite{task4_link}, the aim is to predict the presence or absence and the onset and offset times of sound events. Task 4 provides weakly-labelled data, unlabelled data and simulated strongly-labelled data for training. In Task 5, \textit{Urban sound tagging} \cite{bello2018sonyc}, the aim is to predict urban sound tags of audio recordings recorded in New York. Multiple annotations are provided for each recording, and they do not always agree. Dataset of Task 1, 2 and 5 are weakly labelled. That is, only tags of the audio recording are provided, but not the onset and offset times of the tags. Dataset of Task 3 and 4 are strongly labelled, where the onset and offset of sound events are labelled. 

Recently, neural-network based methods such as fully-connected neural networks \cite{kong2016deep}, convolutional neural networks (CNNs) \cite{choi2016automatic, kong2018dcase, gemmeke2017audio} and recurrent neural networks (RNNs) \cite{cakir2017convolutional} have been used for sound classification and detection, and have achieved state-of-the-art performance in recent DCASE tasks \cite{mesaros2017dcase, mesaros2018multi}. The CNN based methods usually take log mel spectrogram of audio recordings as input and output the presence probability of sound events in either frame-level or clip-level. However, many of these neural network based methods are designed for particular tasks. It is not clear whether a system designed for one task can generalise well to other tasks. 

The motivation of this paper is to investigate the performance of generic systems across several tasks, instead of focusing on particular tasks. In this paper, we investigate cross-task baseline systems based on CNNs starting from our deep neural network (DNN) and CNN based baseline systems for the DCASE 2016 \cite{kong2016deep} and 2018 \cite{kong2018dcase} challenges. In addition to previous work \cite{kong2016deep, kong2018dcase}, we investigate deeper CNN architectures with up to 13 layers, including 5-layer, 9-layer and 13-layer CNNs. We also investigate different pooling strategies after the convolutional layers, including average pooling and max pooling. We compare the performance of different CNN architectures across all five DCASE 2019 challenge tasks. We release the source code of our cross-task systems as baselines for future research. 

This paper is organised as follows. Section 2 introduces the CNN architectures. Section 3 presents experimental results. Section 4 concludes the paper. 

\section{Convolutional neural networks}
CNNs have been widely used in computer vision and have achieved state-of-the-art performance in several tasks such as image classification \cite{simonyan2014very}. A conventional CNN consists of several convolutional layers, where each convolutional layer consists of filters to convolve with the output from the previous convolutional layer. The filters can capture local patterns of the input feature maps, such as edges in lower layers and profiles of objects in higher layers \cite{simonyan2014very}. CNNs have been applied to many audio classification and sound event detection tasks using inputs such as log mel spectrogram \cite{mesaros2016tut, mesaros2018multi}. However, the choice of architecture is usually task-dependent. There is a lack of research investigating the performance of cross-task CNN systems.

We investigate three kinds of CNNs with different depths: a 5-layer CNN, a 9-layer CNN and a 13-layer CNN. The 5-layer CNN is similar to AlexNet \cite{krizhevsky2012imagenet}, which consists of 4 convolutional layers with a kernel size of $ 5 \times 5 $. The later VGG network \cite{simonyan2014very} was proposed to decompose the $ 5 \times 5 $ kernel to a convolutional block consisting of two cascaded convolutional layers with $ 3 \times 3 $ kernels: this inspired the 9-layer and 13-layer CNNs used in this paper with 4 and 6 convolutional blocks. For all the architectures, batch normalisation \cite{ioffe2015batch} is applied after each convolutional operation to speed up and stabilise the training. The ReLU function \cite{simonyan2014very} is used as a non-linearity after batch normalisation. Average pooling or max pooling with a size of $ 2 \times 2 $ is applied after each convolutional block to reduce the feature map size. As the 9-layer CNN perform better than the 5-layer and 9-layer CNNs in experiment, we compare the 9-layer CNN with average pooling named CNN9-avg with the 9-layer CNN with max pooling named CNN9-max. To help the systems robust to frequency shift of sound events, we average out the frequency information in the feature maps of the last convolutional layer. For audio tagging tasks with weakly labelled data, the information over time frames are maxed out, which is designed to select the predominant information over time steps for clip-level classification. Finally, a fully connected layer is applied to predict the presence of sound events either at the clip-level or frame-level. Table \ref{table:CNN_architectures} summaries the CNN architectures used in this paper. For example, $ 5 \times 5 \ @ \ 64 $ indicates a convolutional layer with a kernel size of $ 5 \times 5 $ and an output feature maps number of 64. The parameters number (PN) of models are shown at the bottom of each column in Table \ref{table:CNN_architectures}. For classification tasks, softmax nonlinearity is applied and cross entropy (CE) loss $ l_{\text{CE}} $ is used for training the network:
\begin{equation} \label{eq:cross_entropy}
l_{\text{CE}}(\mathbf{p}, \mathbf{y}) = \sum_{k=1}^{K} y_{k} \text{ln}p_{k}
\end{equation}
\noindent where $ \mathbf{y} = (y_{1}, ..., y_{K}) \in \{ 0, 1 \}^{K} $ is the clip-level or frame-level target and $ K $ is the number of sound classes. The prediction $ \mathbf{p} = (p_{1}, ..., p_{K}) \in [0, 1]^{K} $ is the predicted probability of sound classes. 
For audio tagging and polyphonic sound event detection (SED) tasks, sigmoid nonlinearity is applied. Binary cross entropy (BCE) loss $ l_{\text{BCE}} $ is used for training the network together with the sigmoid output:
\begin{equation} \label{eq:binary_cross_entropy}
l_{\text{BCE}}(\mathbf{p}, \mathbf{y}) = \sum_{k=1}^{K} \left [y_{k} \text{ln}p_{k} + (1-y_{k}) \text{ln} (1 - p_{k})  \right ]. 
\end{equation}
\noindent For the localisation task (Task 3), the target is to predict both the occurrence of sound events and their corresponding azimuth and elevation. We denote the azimuth and elevation target of the $ k $-th sound event as $ y_{k}^{(\text{azi})} $ and $ y_{k}^{(\text{ele})} $. Similarly, the azimuth and elevation of output is denoted as $ o_{k}^{(\text{azi})} $ and $ o_{k}^{(\text{ele})} $. The detection and localisation network is jointly trained using the following loss function:
\begin{equation} \label{eq:seld_loss}
\text{loss} = l_{\text{BCE}}(\mathbf{p}, \mathbf{k}) + \lambda \sum_{k=1}^{K} y_{k} \left (\left |o_{k}^{(\text{azi})} - y_{k}^{(\text{azi})}  \right | + \left |o_{k}^{(\text{ele})} - y_{k}^{(\text{ele})} \right | \right )
\end{equation}
where $ \lambda $ is a weight to balance the SED loss and the localisation loss. We only calculate the MAE between prediction and target in the time steps where target $ y_{k} $ is true for each sound class. 

\begin{table}[]
\caption{CNN architectures. }
\label{table:cnn_architectures}
\resizebox{\columnwidth}{!}{%
\begin{tabular}{ccc}
\hline
\multicolumn{1}{c|}{CNN5} & \multicolumn{1}{c|}{CNN9} & CNN13 \\ \hline
\multicolumn{3}{c}{Log-mel spectrogram} \\ \hline
\multicolumn{1}{c|}{$ \begin{pmatrix} 5 \times 5 \ @ \ 128 \\ \text{BN, ReLU} \end{pmatrix} $} & \multicolumn{1}{c|}{$ \begin{pmatrix} 3 \times 3 \ @ \ 64 \\ \text{BN, ReLU} \end{pmatrix} \times 2 $} & $ \begin{pmatrix} 3 \times 3 \ @ \ 64 \\ \text{BN, ReLU} \end{pmatrix} \times 2 $ \\ \hline
\multicolumn{3}{c}{$ 2 \times 2 \ \text{Pooling} $} \\ \hline
\multicolumn{1}{c|}{$ \begin{pmatrix} 5 \times 5 \ @ \ 128 \\ \text{BN, ReLU} \end{pmatrix} $} & \multicolumn{1}{c|}{$ \begin{pmatrix} 3 \times 3 \ @ \ 128 \\ \text{BN, ReLU} \end{pmatrix} \times 2 $} & $ \begin{pmatrix} 3 \times 3 \ @ \ 128 \\ \text{BN, ReLU} \end{pmatrix} \times 2 $ \\ \hline
\multicolumn{3}{c}{$ 2 \times 2 \ \text{Pooling} $} \\ \hline
\multicolumn{1}{c|}{$ \begin{pmatrix} 5 \times 5 \ @ \ 256 \\ \text{BN, ReLU} \end{pmatrix} $} & \multicolumn{1}{c|}{$ \begin{pmatrix} 3 \times 3 \ @ \ 256 \\ \text{BN, ReLU} \end{pmatrix} \times 2 $} & $ \begin{pmatrix} 3 \times 3 \ @ \ 256 \\ \text{BN, ReLU} \end{pmatrix} \times 2 $ \\ \hline
\multicolumn{3}{c}{$ 2 \times 2 \ \text{Pooling} $} \\ \hline
\multicolumn{1}{c|}{$ \begin{pmatrix} 5 \times 5 \ @ \ 512 \\ \text{BN, ReLU} \end{pmatrix} $} & \multicolumn{1}{c|}{$ \begin{pmatrix} 3 \times 3 \ @ \ 512 \\ \text{BN, ReLU} \end{pmatrix} \times 2 $} & $ \begin{pmatrix} 3 \times 3 \ @ \ 512 \\ \text{BN, ReLU} \end{pmatrix} \times 2 $ \\ \hline
\multicolumn{1}{c|}{$\text{PN:} \ 4,304,320 $} & \multicolumn{1}{c|}{$\text{PN:} \ 4,686,144 $} & \multicolumn{1}{c|}{$ 2 \times 2 \ \text{Pooling} $} \\ \hline
\multicolumn{2}{c|}{\multirow{4}{*}{}} & \multicolumn{1}{c}{$ \begin{pmatrix} 3 \times 3 \ @ \ 1024 \\ \text{BN, ReLU} \end{pmatrix} \times 2 $} \\ \cline{3-3} 
\multicolumn{2}{c|}{} & \multicolumn{1}{c}{$ 2 \times 2 \ \text{Pooling} $} \\ \cline{3-3} 
\multicolumn{2}{c|}{} & \multicolumn{1}{c}{$ \begin{pmatrix} 3 \times 3 \ @ \ 2048 \\ \text{BN, ReLU} \end{pmatrix} \times 2 $} \\ \cline{3-3} 
\multicolumn{2}{c|}{} & \multicolumn{1}{c}{$\text{PN:} \  75,477,312$} \\ \cline{3-3}
\end{tabular}}
\end{table}

\begin{table*}[t]
\centering
\caption{Statistics of DCASE 2019 challenge tasks. }
\label{table:description}
\begin{tabular}{lllllll|l}
 \toprule
 & Task type & Duration & Classes num. & Channels & Sample rate & Label type & Loss \\
 \midrule
 Task 1 & Classification & \textasciitilde 47.2 h & 10 & 1 or 2 & 44.1 kHz or 48 kHz & Weak & CE (clip) \\
 Task 2 & Tagging & \textasciitilde 90.5 h & 80 & 1 & 44.1 kHz & Weak & BCE (clip) \\
 Task 3 & SED + DOA & \textasciitilde 6.7 h & 11 & 4 & 48 kHz & Strong & BCE + MSE (frame) \\
 Task 4 & SED & \textasciitilde 53.3 h & 10 & 1 or 2 & 44.1 kHz & Weak, strong & BCE (clip, frame) \\
 Task 5 & Tagging & \textasciitilde 7.8 h & 29 & 1 & 44.1 kHz & Weak & BCE (clip) \\
 \bottomrule
\end{tabular}
\end{table*}

\section{Experiments}
We apply the described CNNs to the DCASE 2019 tasks. Table \ref{table:description} shows the statistics of the DCASE 2019 tasks. Audio recordings from all tasks are resampled to 32 kHz which contains most energy. Task 1, 2, 4, 5 are trained with clip-level weak labels. Task 3 and the subset of Task 4 with synthetic data are trained with frame-level strong labels. A short-time Fourier transform (STFT) with a Hanning window size of 1024 samples and a hop size of 500 samples is used to extract the spectrogram, so that there are 64 frames in one second. The number 64 is chosen because it is divisible by the power of two which is convenient for the pooling operations in CNNs. Mel filter banks with 64 bins and cut-off frequencies of 50 Hz to 14 kHz are applied on the spectrogram. Then, a logarithm operation is applied to obtain the log-mel spectrogram. The baseline systems for the five tasks in the DCASE 2019 challenge are implemented with Python and PyTorch. The source code is released on github\footnote{https://github.com/qiuqiangkong/dcase2019\textunderscore task1}\footnote{https://github.com/qiuqiangkong/dcase2019\textunderscore task2}\footnote{https://github.com/qiuqiangkong/dcase2019\textunderscore task3}\footnote{https://github.com/qiuqiangkong/dcase2019\textunderscore task4}\footnote{https://github.com/qiuqiangkong/dcase2019\textunderscore task5}. 

\begin{table*}
  \label{table:task1}
  \caption{Development accuracy of Task 1. }
  \vspace{6pt}
  \label{tab:result}
  \centering
  \begin{tabular}{lcccccccc}
    \toprule
    & \multicolumn{1}{c}{\textbf{\textsc{Subtask A}}} & \multicolumn{4}{c}{\textbf{\textsc{Subtask B}}} & \multicolumn{3}{c}{\textbf{\textsc{Subtask C}}} \\
	\cmidrule(lr){2-2} \cmidrule(lr){3 - 6} \cmidrule(lr){7 - 9} 
    Model & Dev. A & Dev. B & Dev. C & Avg. (B,C) & Dev. A & Dev. A & Unknown & Avg.  \\
    \midrule
    Baseline \cite{mesaros2018multi} & 0.625 & 0.396 & 0.431 & 0.414 & 0.619 & 0.542 & 0.431 & 0.487 \\
    CNN5-avg & 0.698 & 0.456 & 0.522 & 0.489 & \textbf{0.691} & 0.581 & 0.481 & \textbf{0.531} \\
    CNN9-avg & \textbf{0.703} & \textbf{0.533} & 0.548 & 0.541 & 0.686 & 0.596 & 0.432 & 0.514 \\ 
    CNN9-max & 0.686 & 0.526 & \textbf{0.567} & \textbf{0.547} & 0.673 & 0.569 & \textbf{0.487} & 0.528 \\
    CNN13-avg & 0.614 & 0.448 & 0.498 & 0.473 & 0.658 & \textbf{0.616} & 0.249 & 0.433 \\
	\bottomrule
\end{tabular}
\end{table*}

\subsection{Task 1, Acoustic scene classification}
Acoustic scene classification \cite{mesaros2018multi} is a task to classify a test recording into one of the provided predefined classes that characterises the environment in which it was recorded. There are 10 sound classes recorded in 12 European countries such as ``airport'' and ``metro station''. This task includes three subtasks: 1) ASC with matched devices, where the testing data are recorded from the same device as the training data, 2) ASC with mismatched recording devices, where the testing data are recorded using different devices as the training data, and 3) open set ASC, where part of the testing data is not encountered in the training data. For each subtask, the evaluation criterion is classification accuracy, which is obtained by averaging the class-wise accuracy of all sound classes. 

Table \ref{table:task1} shows the performance of different CNNs. In Subtask A, the CNN9-avg achieves an accuracy of 0.703, outperforming the 5-layer and 13-layer CNN. In Subtask B, the CNN9-avg and CNN9-max achieve accuracy of 0.541 and 0.547 in the mismatching devices, outperforming the 5-layer and 13-layer CNN. This indicates that the 9-layer CNNs have good performance in classifying mismatching devices. In Subtask C, the 5-layer CNN achieves an average accuracy of 0.531, outperforming the 9-layer and 13-layer CNN.

\subsection{Task 2, Audio tagging with noisy labels and minimal supervision}
Audio tagging with noisy labels and minimal supervision \cite{task2_link} is a task to predict multiple labels of audio clips. The dataset consists of: 1) a small set of curated data from the Freesound Datasets (FSD) \cite{fonseca2017freesound}, which are manually verified and have variable duration from 0.3 s to 30 s; 2) a large set of noisy data from the Yahoo Flickr Creative Commons 100M dataset (YFCC) \cite{fonseca2019learning}, which are not manually verified and may contain incorrect labels. Following \cite{task2_link}, label-weighted label-ranking average precision (lwLRAP) is used for evaluating the performance of the designed systems. 

As there is no official validation set provided, we split the development data to four folds for validation. The systems are trained on fold 2, 3 and 4 and validated on fold 1. Long audio recordings are split into 5-second segments. Short audio recordings are padded to 5-second segments with repeated pattern. Each segment inherit the tag of the audio recording. In Table \ref{table:task2}, the capital letter A, B indicates training on curated FSD and noisy YFCC subset, respectively. A + B indicates training on both FSD and YFCC dataset. We evaluate the lwLRAP on the curated and noisy subset. Table \ref{table:task2} shows that the CNN9-avg trained with A + B achieves lwLRAP of 0.714 and 0.635 on curated and noisy data, outperforming the 5-layer, 13-layer CNN architectures. 

\begin{table}
  \label{table:task2}
  \caption{Development lwLRAP of Task 2.}
  \vspace{6pt}
  \centering
  \begin{tabular}{l p{0.7cm}p{0.7cm}p{0.7cm}p{0.7cm}p{0.7cm}p{0.7cm}}
    \toprule
    & \multicolumn{2}{c}{\textbf{\textsc{A}}} & \multicolumn{2}{c}{\textbf{\textsc{B}}} & \multicolumn{2}{c}{\textbf{\textsc{A + B}}} \\
	\cmidrule(lr){2-3} \cmidrule(lr){4 - 5} \cmidrule(lr){6 - 7} 
    Model & Curated & Noisy & Curated & Noisy & Curated & Noisy \\
    \midrule
    CNN5-avg & 0.806 & 0.246 & 0.393 & 0.613 & 0.669 & 0.616 \\
    CNN9-avg & \textbf{0.822} & \textbf{0.261} & \textbf{0.420} & \textbf{0.629} & \textbf{0.714} & \textbf{0.635} \\ 
    CNN9-max & 0.815 & 0.256 & 0.401 & 0.620 & 0.707 & 0.629 \\
    CNN13-avg & 0.768 & 0.215 & 0.387 & 0.592 & 0.686 & 0.601 \\
	\bottomrule
\end{tabular}
\end{table}

\subsection{Task 3, Sound event localization and detection}
Sound event localisation and detection (SELD) \cite{adavanne2018sound} is a task to detect the onset and offset time of sound events and their directions-of-arrival (DOAs) in azimuth and elevation angles. The development set provides 400 1-minute audio recordings. There are two formats of recording including 4-channel first-order-ambisonic (FOA) and four-channel directional microphone recordings from a tetrahedral array. The impulse responses are collected from five indoor locations at 504 unique combinations of azimuth-elevation-distance. The isolated sound events are from DCASE 2016 Task 2. The F-score and error rate are calculated in one-second segments for evaluating SED. DOA error and frame recall are used for evaluating localisation. 

We extract the log-mel spectrogram of the 4-channel FOA as input to the CNNs. Different from the baseline system \cite{adavanne2018sound} using additional phase information as input, we only use the magnitude of log-mel spectrogram as input. In training, audio recordings are split into overlapped 10-second segments. The loss function in equation (\ref{eq:seld_loss}) is used for training. The evaluation follows \cite{adavanne2018sound}. Table \ref{table:task3} shows the performance of the CNN systems. The 5-layer and 9-layer CNNs achieve similar results in SED and localisation, outperforming the 13-layer CNN. A DOA error of 42.8$^\circ$ using the 13-layer CNN. 
\begin{table}[t]
\centering
\caption{Development performance of Task 3. }
\label{table:task3}
\resizebox{\columnwidth}{!}{%
\begin{tabular}{l p{0.9cm}p{0.9cm}p{0.9cm}p{0.9cm}p{0.9cm}}
 \toprule
 & Error rate & F1 score & DOA error & Frame recall & SELD score \\
 \midrule
 Baseline \cite{adavanne2018sound} & 0.34 & 79.9\% & \textbf{28.5$^\circ$} & \textbf{85.4\%} & - \\
 CNN5-avg & 0.33 & \textbf{80.7\%} & 54.4$^\circ$ & 77.0\% & 0.263 \\
 CNN9-avg & \textbf{0.32} & 80.5\% & \ang{44.0} & 77.1\% & \textbf{0.248} \\
 CNN9-max & 0.34 & 79.4\% & 45.6$^\circ$ & 76.3\% & 0.260 \\
 CNN13-avg & 0.42 & 72.8\% & 42.8$^\circ$ & 71.4\% & 0.303 \\
 \bottomrule
\end{tabular}}
\end{table}
\subsection{Task 4, Sound event detection in domestic environments}
Sound event detection in domestic environments \cite{task4_link} is a task to detect the onset and offset time steps of sound events in domestic environments. The datasets are from AudioSet \cite{gemmeke2017audio}, FSD \cite{fonseca2017freesound} and SINS dataset \cite{dekkers2017sins}. The aim of this task is to investigate whether real but weakly annotated data or synthetic data is sufficient for designing SED systems. There are 1578 real audio recordings with weak labels, 2045 synthetic recordings with strong labels, and 14412 unlabelled in-domain recordings in the dataset. Audio recordings are 10 seconds in duration and consist of polyphonic sound events from 10 sound classes. 

Table \ref{table:task4} shows the performance of CNN systems using the real weakly-labelled subset for training. CNN9-max achieves an event-based F1 of 24.1\%, a segment-based F1 of 63.0\% and an audio tagging mAP of 0.791, outperforming the other CNN systems. Different from other tasks, $ 2 \times 2 $ max pooling outperforms $ 2 \times 2 $ average pooling in this task. In addition, we observe that using only the real weakly labelled data for training achieves an event-based F1 of 24.1\%, outperforming the result of using synthetic weakly and strongly labelled data for training of 13.0\% and 13.4\%, respectively. 
\begin{table}[t]
\centering
\caption{Development performance of Task 4. }
\label{table:task4}
\begin{tabular}{l p{1.8cm}p{1.7cm}p{1.7cm}}
 \toprule
 & Event-based F1 & segment-based F1 & Audio tagging mAP\\
 \midrule
 Baseline \cite{task4_link} & 23.5\% & 54.7\% & - \\
 CNN5-avg & 18.0\% & 59.9\% & 0.765 \\
 CNN9-avg & 20.0\% & 58.6\% & 0.778 \\
 CNN9-max & \textbf{24.1}\% & \textbf{63.0}\% & \textbf{0.791} \\
 CNN13-avg & 17.0\% & 58.3\% & 0.740 \\
 \bottomrule
\end{tabular}
\end{table}
\subsection{Task 5, Urban sound tagging}
Urban sound tagging \cite{bello2018sonyc} is a task to predict the presence or absence of different type of sounds recorded in New York City. This task has established a set of coarse-grained and fine-grained classes. The coarse-grained and fine-grained tags consist of 7 and 23 sound classes under a hierarchical taxonomy. There are 2351 10-second audio recordings for development. Each audio recording is annotated by one or more annotators. The label of a single audio recording can be different depending on the annotator. We apply an OR function to aggregate the labels from different annotators. That is, an audio recording for a sound class is positive if at least one annotator labels it as positive. Different from the baseline system \cite{bello2018sonyc} trained on AudioSet features \cite{gemmeke2017audio}, we train the CNNs on the log mel spectrogram from scratch. Table \ref{table:task5} shows that CNN9-avg achieves fine-grained and coarse-grained micro AUPRCs of 0.672 and 0.782, outperforming the 5-layer, 13-layer CNNs.
\begin{table}
  \caption{Development performance of Task 5. }
  \vspace{6pt}
  \label{table:task5}
  \centering
  \resizebox{\columnwidth}{!}{%
  \begin{tabular}{l p{0.8cm}p{0.8cm}p{0.8cm}p{0.8cm}p{0.8cm}p{0.8cm}}
    \toprule
    & \multicolumn{3}{c}{\textbf{\textsc{Fine-grained}}} & \multicolumn{3}{c}{\textbf{\textsc{Coarse-grained}}} \\
	\cmidrule(lr){2-4} \cmidrule(lr){5-7} 
	& \small{Micro AUPRC} & \small{Micro F1} & \small{Macro AUPRC} & \small{Micro AUPRC} & \small{Micro F1} & \small{Macro AUPRC} \\
	\midrule
 Baseline & 0.672 & 0.502 & 0.427 & 0.742 & 0.507 & 0.530 \\
 CNN5-avg & 0.659 & \textbf{0.395} & \textbf{0.451} & 0.765 & \textbf{0.552} & 0.570 \\
 CNN9-avg & \textbf{0.672} & 0.371 & 0.433 & \textbf{0.782} & 0.519 & \textbf{0.628} \\
 CNN9-max & 0.668 & 0.337 & 0.457 & 0.765 & 0.493 & 0.555 \\
 CNN13-avg & 0.637 & 0.368 & 0.428 & 0.742 & 0.491 & 0.333 \\

	\bottomrule
\end{tabular}}
\end{table}
\begin{figure}[t]
  \centering
  \centerline{\includegraphics[width=\columnwidth]{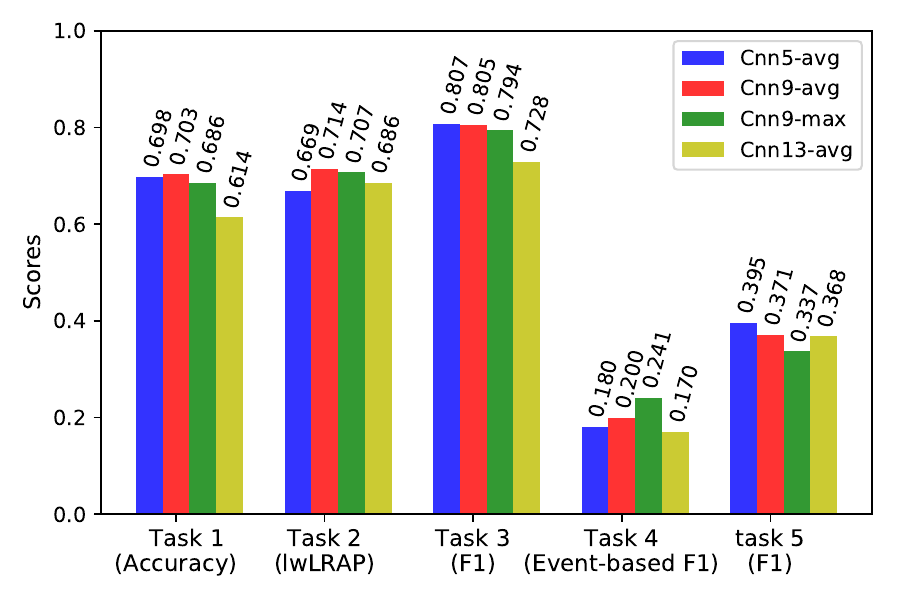}}
  \caption{Performance of cross-task CNN systems on the DCASE 2019 challenge Task 1 - 5.}
  \label{fig:results}
\end{figure}

\section{Conclusion}
This paper presents cross-task convolutional neural network baseline systems for the DCASE 2019 tasks. We investigated the performance of 5-layer, 9-layer and 13-layer CNN systems with average or max pooling on five tasks. Figure \ref{fig:results} shows the performance of cross-task CNN systems on the DCASE 2019 challenge Task 1 - 5. Although different tasks apply different evaluation metrics, it can be seen that each individual task may favor particular CNN systems. Average pooling achieves better performance than max pooling in 4 out of 5 tasks. The 9-layer CNNs achieve better performance than the 5-layer and 13-layer CNNs in 3 out of 5 tasks. Overall, the 9-layer CNN with average pooling was shown to perform good for a majority of tasks. We release the cross-task baseline code. In future, we will continue to explore the cross-task structures with more neural network architectures such as recurrent neural networks.

\section{Acknowledgement}
This research was supported by EPSRC grant EP/N014111/1 ``Making Sense of Sounds'' and a Research Scholarship from the China Scholarship Council (CSC) No. 201406150082. 

% -------------------------------------------------------------------------
% Either list references using the bibliography style file IEEEtran.bst
\bibliographystyle{IEEEtran}
\bibliography{refs}
%
% or list them by yourself
% \begin{thebibliography}{9}
% 
% \bibitem{waspaa19web}
%   \url{http://www.waspaa.com}.
%
% \bibitem{IEEEPDFSpec}
%   {PDF} specification for {IEEE} {X}plore$^{\textregistered}$,
%   \url{http://www.ieee.org/portal/cms_docs/pubs/confstandards/pdfs/IEEE-PDF-SpecV401.pdf}.
%
% \bibitem{PDFOpenSourceTools}
%   Creating high resolution {PDF} files for book production with 
%   open source tools, 
%   \url{http://www.grassbook.org/neteler/highres_pdf.html}.
%
% \bibitem{eWilliams1999}
% E. Williams, \emph{Fourier Acoustics: Sound Radiation and Nearfield Acoustic
%   Holography}. London, UK: Academic Press, 1999.
% 
% \bibitem{ieeecopyright}
%   \url{http://www.ieee.org/web/publications/rights/copyrightmain.html}.
%
% \bibitem{cJones2003}
% C. Jones, A. Smith, and E. Roberts, ``A sample paper in conference
%   proceedings,'' in \emph{Proc. IEEE ICASSP}, vol. II, 2003, pp. 803--806.
% 
% \bibitem{aSmith2000}
% A. Smith, C. Jones, and E. Roberts, ``A sample paper in journals,'' 
%   \emph{IEEE Trans. Signal Process.}, vol. 62, pp. 291--294, Jan. 2000.
% 
% \end{thebibliography}

\end{sloppy}
\end{document}